\documentclass[twocolumn,showpacs,floatfix,superscriptaddress,amsmath,amssymb,prl]{revtex4}
\usepackage{amsfonts}
\usepackage{amsmath}
\usepackage{amssymb}
\usepackage{graphicx}
\usepackage{hyperref}
\usepackage{ulem}

\newcommand{\be}{\begin{equation}}
\newcommand{\ee}{\end{equation}}
\newcommand{\bra}[1]{\mbox{$\langle #1 |$}}
\newcommand{\ket}[1]{\mbox{$| #1 \rangle$}}
\newcommand{\braket}[2]{\mbox{$\langle #1  | #2 \rangle$}}

\begin{document}

\title{Ground state fidelity from tensor network representations}

\author{Huan-Qiang Zhou}
\affiliation{Centre for Modern Physics and Department of Physics,
Chongqing University, Chongqing 400044, The People's Republic of
China}

\author{Roman Or\'us}
\affiliation{School of Physical Sciences, University of Queensland,
Brisbane, Qld 4072, Australia}

\author{Guifre Vidal}
\affiliation{School of Physical Sciences, University of Queensland,
Brisbane, Qld 4072, Australia}

\begin{abstract}
For any $D$-dimensional quantum lattice system, the fidelity between
two ground state many-body wave functions is mapped onto the
partition function of a $D$-dimensional classical statistical vertex
lattice model with the same lattice geometry. The fidelity per
lattice site, analogous to the free energy per site, is well defined
in the thermodynamic limit and can be used to characterize the phase
diagram of the model. We explain how to compute the fidelity per
site in the context of tensor network algorithms, and demonstrate
the approach by analyzing the two-dimensional quantum Ising model
with transverse and parallel magnetic fields.

\end{abstract}


\maketitle

The discoveries of high-$T_c$ superconductors and fractional quantum
Hall liquids have stimulated a surge of activities in the study of
quantum phase transitions (QPTs)~\cite{sachdev}. The conventional
description of QPTs in condensed matter physics is in terms of
orders and fluctuations. The Landau-Ginzburg-Wilson paradigm
describes symmetry-breaking orders quantified by a local order
parameter, whose non-zero value characterizes a symmetry-broken
phase.  Continuous QPTs beyond the Landau-Ginzburg-Wilson
paradigm also exist. They are described in terms of the so-called
topological/quantum orders~\cite{wen} and are relevant to emergent phenomena
in strongly correlated electron systems, with nonlocal order
parameters as a salient feature.

By using concepts of quantum information science, recently two new
approaches to study QPTs have been proposed. They focus on
properties of ground state wave functions of the quantum many-body
system, namely \textit{entanglement}
~\cite{preskill,osborne,vidal,korepin,levin,entanglement,review} and
\textit{fidelity} ~\cite{zanardi,zhou,fidelity}, and turn out to be
very successful at detecting quantum critical behaviors. In
particular, the entanglement entropy is exploited to reveal
qualitatively different behaviors at and off quantum criticality
~\cite{vidal,korepin}, whereas the fidelity, a measure of
distinguishability of states in the system's Hilbert space, is shown
to be able to capture drastic changes in quantum ground states when
the system undergoes a QPT, regardless of what type of internal
order is present ~\cite{zhou}. Both approaches have been shown to be
insightful in the context of already well-understood systems, but in
practice, when applied to a generic system, they still rely on our
ability to compute certain properties of ground state wave
functions, which is in general a very difficult task.

On the other hand, significant progress has also been made recently
in the classical simulation of quantum many-body systems by using a
\textit{tensor network} (TN) to represent the wave function.
Examples of TNs include a matrix product state (MPS)
\cite{MPS,TEBD,TI_MPS} for systems in one spatial dimension and the
projected entangled-pair state (PEPS) \cite{PEPS} in two and higher
spatial dimensions. For systems invariant under translations,
particularly efficient algorithms have been proposed to compute the
ground state for infinite systems, both in one \cite{iTEBD} and two
\cite{iPEPS} spatial dimensions, as well as for finite systems with
periodic boundary conditions (PBC) \cite{PBC_TN}.

The purpose of this Letter is two-fold. We consider a system, either
infinite or finite with PBC, defined on a $D$-dimensional lattice
and such that its ground state is invariant under translations
\cite{TI}. First, we show that the fidelity between two ground
states can be mapped onto the partition function of a
$D$-dimensional classical statistical vertex lattice model with the
same lattice geometry. This is achieved by exploiting the fact that
the two ground states can be represented in terms of a TN where all
the tensors are copies of one single tensor. The fidelity per
lattice site, introduced in Ref.~\cite{zhou}, is naturally
interpreted as the free energy per site of this $D$-dimensional
classical statistical vertex lattice model, and as such it is well
defined in the thermodynamic limit (even though the fidelity itself
becomes zero). Second, we consider the practical computation of the
fidelity per lattice site, both for finite and infinite systems,
within the framework of TN algorithms for translationally invariant
systems \cite{iTEBD,iPEPS,PBC_TN}. As a result, we obtain a viable
scheme to determine the ground state phase diagram of a system
without prior knowledge of order parameters. We demonstrate the
approach by analysing the two-dimensional quantum Ising model with
both transverse and parallel magnetic fields. First and second order
phase transitions, as well as stable fixed points, are clearly
identified.

{\it Generalities.} Consider a finite quantum lattice system $S$ in
$D$ dimensions described by a Hamiltonian $H(\lambda)$, where
$\lambda$ is a control parameter~\cite{multiple}. For two ground
states $|\psi(\lambda_1)\rangle$ and $|\psi(\lambda_2)\rangle$
corresponding to two different values $\lambda_1$ and $\lambda_2$ of
the control parameter $\lambda$, the ground state fidelity
$F(\lambda_1,\lambda_2) = | \langle
\psi(\lambda_2)|\psi(\lambda_1)\rangle|$ asymptotically scales as
$F(\lambda_1,\lambda_2) \sim {d(\lambda_1,\lambda_2)}^N$, with $N$
the number of sites in the lattice. Here, $d(\lambda_1,\lambda_2)$
is the scaling parameter, introduced in Ref.~\cite{zhou} for
one-dimensional quantum systems, which characterizes how fast the
fidelity goes to zero when the thermodynamic limit is approached.
Physically, the scaling parameter $d(\lambda_1,\lambda_2)$ is the
{\it averaged} fidelity per lattice site,
\begin{equation}
\ln d(\lambda_1,\lambda_2) \equiv \lim_{N \rightarrow \infty} \frac {\ln
F(\lambda_1,\lambda_2)}{N}, \label{d}
\end{equation}
which is seen to be well defined in the thermodynamic limit even if
$F(\lambda_1,\lambda_2)$ becomes trivially zero. It satisfies the
properties inherited from the fidelity $F(\lambda_1,\lambda_2)$: (i)
normalization $d(\lambda,\lambda)=1$; (ii) symmetry
$d(\lambda_1,\lambda_2)=d(\lambda_2,\lambda_1)$; and (iii) range $0
\le d(\lambda_1,\lambda_2)\le 1$. Additionally, in a finite system
we can define a finite-size analogue of $d(\lambda_1,\lambda_2)$
through
\begin{equation}
\ln d_N(\lambda_1,\lambda_2) \equiv \frac {\ln
F(\lambda_1,\lambda_2)}{N}. \label{dN}
\end{equation}

As argued in Ref.~\cite{zhou}, the fidelity per lattice site
$d(\lambda_1,\lambda_2)$ succeeds in capturing nontrivial
information including stable and unstable fixed points along
renormalization group flows. Specifically, suppose the system $S$
undergoes a QPT at a transition point $\lambda_c$. Then
$d(\lambda_1,\lambda_2)$ exhibits singular behaviors when
$\lambda_1$ crosses $\lambda_c$ for a fixed $\lambda_2$, or
$\lambda_2$ crosses $\lambda_c$ for a fixed $\lambda_1$. That is, a
transition point $\lambda_c$ is characterized as a \textit{pinch
point} $(\lambda_c,\lambda_c)$ for \textit{continuous} QPTs: the
intersection of two singular lines $\lambda_1=\lambda_c$ and
$\lambda_2=\lambda_c$ on the two-dimensional surface defined by
$d(\lambda_1,\lambda_2)$ as a function of $\lambda_1$ and
$\lambda_2$.  For first order QPTs, $d(\lambda_1,\lambda_2)$)
becomes discontinuous (as either $\lambda_1$ or $\lambda_2$ crosses
a transition point)~\cite{firstorder}.

{\it Mapping onto a $D$-dimensional classical statistical vertex
lattice model.} As it is well known, there is a remarkable mapping
from a $D$-dimensional quantum system to an equivalent
$(D+1)$-dimensional classical system with imaginary time as an extra
dimension~\cite{book1,book2}. Here, we discuss another mapping, one
from the ground state fidelity $F(\lambda_1,\lambda_2)$ for a
$D$-dimensional quantum lattice model onto the partition function of
a $D$-dimensional classical statistical vertex lattice model. This
mapping implies that we can take advantage of the whole machinery of
the transfer matrix formulation in statistical mechanics. As we
discuss below, it also means that we can compute the fidelity per
lattice site $d(\lambda_1,\lambda_2)$ by exploiting the TN
algorithms of Refs. \cite{iTEBD,iPEPS,PBC_TN}.

\begin{figure}[hb]
  \begin{center}
    \includegraphics[width=8cm]{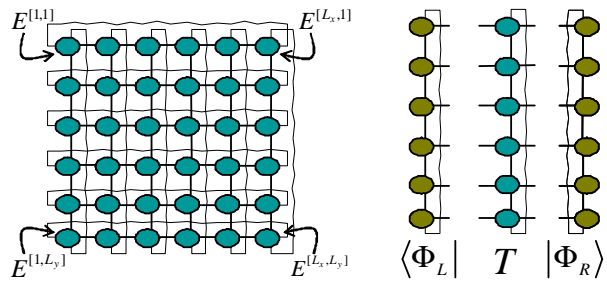}
  \end{center}
  \caption{(color online) Diagrammatical representation of several tensor networks.
  \textit{Left:} two-dimensional tensor network for the ground state fidelity $F(\lambda_1,\lambda_2)$ in a system defined on a torus. \textit{Right:} matrix product operator (MPO) for the corresponding one-dimensional transfer matrix $T(\lambda_1,\lambda_2)$, and matrix product state (MPS) for the left and right eigenvectors of $T$, $|\Phi_L\rangle$ and $|\Phi_R\rangle$, with the largest eigenvalue $\mu$.}
  \label{network}
\end{figure}

To establish this mapping, we recall that \textit{any} state of a
quantum lattice system may be represented in terms of a TN, such as
an MPS for one-dimensional systems or a PEPS for systems in $D\geq
2$ dimensions \cite{TEBD,PEPS}. As a concrete example, let us
consider a square lattice on a torus with $N = L_x \times L_y$
sites, where each site, labeled by a vector $\vec{r} = (x,y)$, is
represented by a $q$-dimensional Hilbert space $V^{[\vec{r}]} \equiv
\mathfrak{C}^q$. A PEPS for a state $|\psi (\lambda) \rangle$
consists of a set of tensors $A^{[\vec{r}]}$, one tensor per lattice
site. Each tensor is made of complex numbers $A^{[\vec{r}]s}_{\alpha
\beta \gamma \delta}$ labeled by one {\it physical} index $s$ and
four {\it bond} indices $\alpha,\beta,\gamma$ and $\delta$ (in a
generic case, there will be one bond index for each outgoing link of
site $\vec{r}$). The physical index $s$ runs over a basis of
$V^{[\vec{r}]}$, so that $s=1, \cdots,q$, whereas each bond index
takes $Q$ values, with $Q$ some inner dimension of bonds in the
valence bond picture, which connects the tensors in the nearest
neighbor sites. In terms of the PEPS representation, the ground
state fidelity turns out to be equivalent to the partition function
of a two-dimensional classical statistical vertex lattice model, see
Fig. (\ref{network}), with the statistical ``weights"
\begin{equation}
E^{[\vec{r}]}_{\tilde{\alpha}\tilde{\beta}\tilde{\gamma}\tilde{\delta}}(\lambda_1,\lambda_2)
\equiv \sum_s \left[A^{[\vec{r}]s}_{\alpha' \beta' \gamma'
\delta'}(\lambda_2)\right]^* A^{[\vec{r}]s}_{\alpha \beta \gamma
\delta}(\lambda_1), \label{eq:weights}
\end{equation}
where the tilded indices are combined pairs of indices:
$\tilde{\alpha} \equiv(\alpha,\alpha')$ and so on. By inspecting
definitions (\ref{d}) and (\ref{dN}), one concludes that the
logarithm of $d_N(\lambda_1,\lambda_2)$ is formally equivalent to
the free energy per site in the two-dimensional classical
statistical vertex lattice model~\cite{classical} (up to an
irrelevant prefactor linear in temperature). This argument is valid
for any lattice geometry in any dimensions~\cite{1Dmapping}.
Therefore, the fact that QPTs may be detected as singularities in
$d(\lambda_1,\lambda_2)$ matches the conventional wisdom that phase
transition points are reflected as singularities, in the
thermodynamic limit, of the free energy for classical systems.

Some remarks are in order. First, the mapping is {\it exact} both
for finite lattices (possibly for a large $Q$) and infinite lattices
(infinite $Q$). Second, for periodic systems that are invariant
under translations, one can always build a TN where all the tensors
are the same (often at the cost of increasing $Q$) by using results
in \cite{TI_MPS,TI_PEPS} and generalizations thereof. Finally, in
practical computations as described below, the exact ground state is
approximated, in a controlled way, by a TN with reasonably small
$Q$.


{\it Fidelity per lattice site from tensor network representations.}
From now on we specialize to a $D$-dimensional lattice system that is invariant under
translations by one lattice site \cite{TI}. We explain how to obtain the fidelity per
site, both in infinite and finite (but large) systems. As a first step, we use the TN
algorithms \cite{iTEBD,iPEPS,PBC_TN} to compute a TN representation for the ground
states $|\psi (\lambda_1) \rangle$ and $|\psi (\lambda_2) \rangle$ in terms of
site-independent tensors $A^{[\vec{r}]}(\lambda_1)$ and $A^{[\vec{r}]}(\lambda_2)$,
that we use to build the (also site-independent) statistical weights
$E^{[\vec{r}]}(\lambda_1,\lambda_2)$. We notice that all these tensors depend on the
lattice size $N$.

The fidelity $F(\lambda_1,\lambda_2)$, regarded as the partition
function of a $D$-dimensional classical statistical vertex lattice
model with weights $E^{[\vec{r}]}(\lambda_1,\lambda_2)$, is the
trace of a power of some \textit{transfer matrix} $T$,
\begin{equation}
    F(\lambda_1,\lambda_2) = {\rm Tr} (T^{L_x}).
\end{equation}
Here $T$, a $(D\!-\!1)$-dimensional tensor network itself, is made of all the tensors $E^{[\vec{r}]}$ contained in some regular slice of the TN for $F(\lambda_1,\lambda_2)$, where the latter consists of exactly $L_x$ identical such slices, see Fig. (\ref{network}). Let $\mu_{\alpha}$ be the eigenvalues of $T$, with $|\mu_{0}|\geq |\mu_{1}| \geq \cdots \geq |\mu_{\alpha_{max}}|$. Then the fidelity reads
\begin{equation}
    F(\lambda_1,\lambda_2) = \sum_{\alpha=0}^{\alpha_{max}} \mu_{\alpha}^{L_x} = \mu_0^{L_x}\left(1 + \sum_{\alpha=1}^{\alpha_{max}} \left(\frac{\mu_{\alpha}}{\mu_{0}}\right)^{L_x}\right) , ~~~~
\end{equation}
so that for large $L_x$, and assuming $|\mu_0|>|\mu_1|$ \cite{non-degenerate},
\begin{equation}
    d_N(\lambda_1,\lambda_2) = \mu_0\left[1 + O\left(\frac{1}{L_x}\left(\frac{\mu_1}{\mu_0}\right)^{L_x}\right)\right].
\end{equation}
That is, $d_N(\lambda_1,\lambda_2)$ is given by the largest
eigenvalue $\mu_0$ of $T$ up to corrections that decay exponentially
in the linear system size $L_x$. Our next task is to determine
$\mu_0$, which in general depends on $N$, $\lambda_1$, and
$\lambda_2$.

First we compute the left and right eigenvectors $|\Phi_L\rangle$ and $|\Phi_R\rangle$ of $T$ corresponding to $\mu_0$,
\begin{equation}
    \langle \Phi_L| ~T = \langle \Phi_L|~ \mu_0,~~~~~~ T~|\Phi_R\rangle = \mu_0 ~|\Phi_R\rangle,
\end{equation}
where we use a $(D-1)$-dimensional TN to represent them. This is
achieved (again with the TN algorithms \cite{iTEBD,iPEPS,PBC_TN}) by
exploiting the fact that, e.g., $|\Phi_R\rangle \sim \lim_{p
\rightarrow \infty} T^{p} \ket{\Psi_0}$ for an arbitrary state
$\ket{\Psi_0}$ such that $\braket{\Psi_0}{\Phi_R}\neq 0$. After
normalizing the states so that $\braket{\Phi_L}{\Phi_R}=1$, we
obtain $\mu_0$ from
\begin{equation}
    \mu_0 = \bra{\Phi_L}T\ket{\Phi_R},
\end{equation}
by evaluating a $(D-1)$-dimensional TN for $\bra{\Phi_L}T\ket{\Phi_R}$, see Fig. (\ref{network}). At this point, we notice that we can use the techniques that we have just discussed in order to evaluate this new TN, by reducing the calculation to a $(D-2)$-dimensional TN, and so forth.

We illustrate the procedure with two simple cases: (i) periodic
chains, $D=1$; (ii) periodic square lattices, $D=2$.

Case (i) \cite{previous}: Each ground state is represented as an MPS
that consists of $N$ copies of the tensor $A_{\alpha \beta}^{s}$,
with one physical index $s$ and two bond indices $\alpha$ and
$\beta$. The zero-dimensional transfer matrix $T$ is given by
$E_{\tilde{\alpha}\tilde{\beta}}(\lambda_1,\lambda_2) \equiv \sum_s
(A^{s}_{\alpha' \beta'}(\lambda_2))^* A_{\alpha
\beta}^{s}(\lambda_1)$, and its diagonalization produces the
eigenvalues $\mu_0,\cdots,\mu_{\alpha_{max}}$.

Case (ii): Each ground state is represented as a PEPS on a torus with $N=L_x \times L_y$ sites, see Fig. (\ref{network}). The one-dimensional transfer matrix $T$ is a matrix product operator (MPO) with tensors given by the statistical weights $E^{[\vec{r}]}$ of Eq. (\ref{eq:weights}). Its left and right eigenvectors $|\Phi_L\rangle$ and $|\Phi_R\rangle$ with maximal eigenvalue $\mu_0$ are represented as MPSs with tensors $L^{s}_{\alpha\beta}$ and $R^{s}_{\alpha\beta}$. The zero-dimensional transfer matrix $T'$ reads
\begin{equation}
    T'_{\tilde{\epsilon}\tilde{\gamma}} = \sum_{s,s'} (L^s_{\epsilon\gamma} )^*E_{\epsilon' s'\gamma' s} R^{s'}_{\epsilon'' \gamma''},
\end{equation}
where $\tilde{\epsilon} = (\epsilon,\epsilon',\epsilon'')$ and $\tilde{\gamma}$ are composite indices. Let $\mu_0'$ be the largest eigenvalue of $T'$. Then, up to corrections that vanish exponentially fast in $L_x$ and $L_y$, we have
\begin{equation}
    F(\lambda_1,\lambda_2) \approx \mu_0^{L_x} = (\mu_0'^{L_y})^{L_x} = (\mu'_0)^N,
\end{equation}
so that
\begin{equation}
    d_N(\lambda_1,\lambda_2) \approx \mu'_0,~~~~~d(\lambda_1,\lambda_2) = \mu'_0
\end{equation}
for the finite and infinite cases, respectively.

\begin{figure}[ht]
  \begin{center}
    \includegraphics[width=9.5cm]{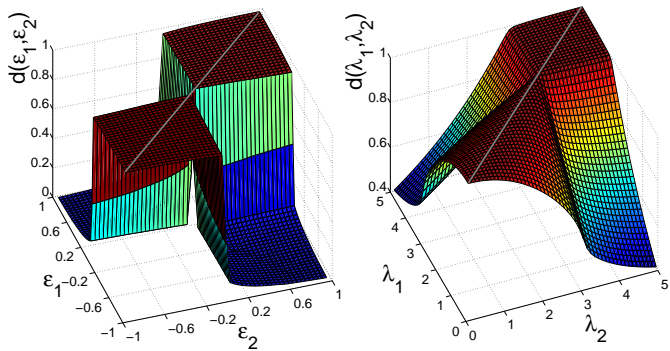}
  \end{center}
  \caption{(color online) Fidelity per lattice site for ground states of the two-dimensional quantum Ising model, Eq. (\ref{HXY}), which is one along the diagonal.
  \textit{Left:} for $\lambda=2.5$ (i.e. $\lambda < \lambda_c$), $d(\epsilon_1,\epsilon_2)$ displays a discontinuity at the lines $\epsilon_1=0$ and $\epsilon_2=0$, which indicates the presence of a first order phase transition. \textit{Right:} for $\epsilon=0$, $d(\lambda_1,\lambda_2)$ has a pinch point at $(\lambda_c,\lambda_c)$, indicating the presence of a second order phase transition.}
  \label{fidelity}
\end{figure}

{\it Example: the two-dimensional quantum Ising model with
transverse and parallel magnetic fields.}  As a test, we compute the
fidelity per lattice site $d(\lambda_1,\lambda_2)$ for the
two-dimensional quantum Ising model in the thermodynamic limit, as
described by the Hamiltonian
\begin{equation}
H= -\sum_{(\vec{r},\vec{r}')}  \sigma^{[\vec{r}]}_z \sigma^{[\vec{r}']}_z  -\lambda \sum_{\vec{r}} \sigma^{[\vec{r}]}_x - \epsilon \sum_{\vec{r}} \sigma^{[\vec{r}]}_z
. \label{HXY}
\end{equation}
Here $\sigma^{[\vec{r}]}_x$ and $\sigma^{[\vec{r}]}_z$ are the Pauli
matrices at the lattice site $\vec{r}$, with the control parameters
$\lambda$ and $\epsilon$ being the transverse and parallel magnetic
fields. For $\epsilon=0$, the system has a second order phase
transition at $\lambda_c \approx 3.044$~\cite{blote}, whereas for
$\lambda<\lambda_c$, a first order phase transition occurs when
$\epsilon$ changes sign. We plot $d(\epsilon_1,\epsilon_2)$ and
$d(\lambda_1,\lambda_2)$ in Fig.~\ref{fidelity}, as computed from
the infinite PEPS algorithm~\cite{iPEPS} with bond dimension $2$. We
can clearly identify the first and second order phase transitions by
a discontinuity in $d(\epsilon_1,\epsilon_2)$ and a pinch point in
$d(\lambda_1,\lambda_2)$, respectively. The two stable fixed points
at $\lambda =0$ and $\lambda =\infty$ are also characterized as the
global minima of $d(\lambda_1,\lambda_2)$.


{\it Summary and outlook.} The fidelity per site
$d(\lambda_1,\lambda_2)$ allows to determine the zero temperature
phase diagram of a quantum lattice system without prior knowledge of
order parameters. Here, we have shown how to compute
$d(\lambda_1,\lambda_2)$ in the context of the TN algorithms of
Refs. \cite{iTEBD,iPEPS,PBC_TN}. We envisage that this approach will
become a preferred strategy to scan a quantum lattice system for
possible phases and phase transitions, perhaps as a first step of a
more comprehensive method that will subsequently characterize each
phase in terms of order parameters, etc. An interesting question is
to see whether or not the scheme works for systems with topological
orders. On the other hand, further work is needed to perform finite
size scaling and extract the correlation length critical exponent by
exploiting the finite TN algorithms, which is currently under
investigation.

We thank L. Masanes for insightful conversations. Support from the
Natural Science Foundation of China and the Australian Research
Council (No. FF0668731) is acknowledged.



\begin{thebibliography}{99}

\bibitem{sachdev} S. Sachdev, \textit{Quantum Phase Transitions},
Cambridge University Press, 1999, Cambridge.

\bibitem{wen} X.-G. Wen, \textit{Quantum Field Theory of Many-Body
Systems}, Oxford University Press, 2004, Oxford.

\bibitem{preskill} J. Preskill, J. Mod. Opt. \textbf{47}, 127
(2000).

\bibitem{osborne} T.J. Osborne and M.A. Nielsen, Phys. Rev. A
\textbf{66}, 032110 (2002); A. Osterloh \textit{et al.}, Nature
\textbf{416}, 608 (2002).

\bibitem{vidal} G. Vidal \textit{et al.}, Phys. Rev. Lett.
\textbf{90}, 227902 (2003); G. Vidal, arXiv:cond-mat/0512165; G.
Evenbly and G. Vidal, arXiv:0710.0692.

\bibitem{korepin} V.E. Korepin, Phys. Rev. Lett. 92, 096402 (2004); G.C. Levine, Phys.
Rev. Lett. 93, 266402 (2004); G. Refael and J.E. Moore, Phys. Rev.
Lett. 93, 260602 (2004); P. Calabrese and J. Cardy, J. Stat. Mech.
P06002 (2004).

\bibitem{levin} A. Kitaev and J.
Preskill, hep-th/0510092; M. Levin and X.-G. Wen, cond-mat/0510613.

\bibitem{entanglement} F. Verstraete, M.A. Martin-Delgado, and J.I. Cirac, Phys. Rev. Lett.
92, 087201 (2004); W. D¨ur \textit{et al.}, Phys. Rev. Lett. 94,
097203 (2005); H. Barnum \textit{et al.}, Phys. Rev. Lett. 92,
107902 (2004).

\bibitem{review} L. Amico \textit{et al.}, quant-ph/0703044.

\bibitem{zanardi} P. Zanardi and N. Paunkovi\'{c},
Phys. Rev. E \textbf{74}, 031123 (2006).

\bibitem{zhou} H.-Q. Zhou and J.P. Barjaktarevi$\check{\rm c}$, cond-mat/0701608;
H.-Q. Zhou, J.-H. Zhao, and B. Li, arXiv:0704.2940; H.-Q. Zhou,
arXiv:0704.2945.

\bibitem{fidelity} P. Zanardi, M. Cozzini, and P. Giorda, cond-mat/ 0606130; N. Oelkers
and J. Links, Phys. Rev. B \textbf{75}, 115119 (2007); M. Cozzini,
R. Ionicioiu, and P. Zanardi, cond-mat/0611727; L. Campos Venuti and
P. Zanardi, Phys. Rev. Lett. \textbf{99}, 095701 (2007); P.
Buonsante and A. Vezzani, Phys. Rev. Lett. \textbf{98}, 110601
(2007); W.-L. You, Y.-W. Li, and S.-J. Gu, Phys. Rev. E \textbf{76},
022101 (2007); S.J. Gu \textit{et al.}, arXiv:0706.2495; M.F. Yang,
arXiv:0707.4574; Y.C. Tzeng and M.F. Yang, arXiv:0709.1518.

\bibitem{MPS} M. Fannes, B. Nachtergaele, and R. F. Werner,
Comm. Math. Phys. \textbf{144}, 443 (1992); J. Funct. Anal.
\textbf{120}, 511 (1994); S. \"Ostlund and S. Rommer, Phys. Rev.
Lett. \textbf{75}, 3537 (1995).

\bibitem{TEBD} G. Vidal, Phys. Rev. Lett. \textbf{91}, 147902 (2003); G. Vidal, Phys. Rev. Lett. 93, 040502 (2004).


\bibitem{TI_MPS} D. Perez-Garcia \textit{et al.}, Quantum Inf. Comput. 7, 401 (2007),
arXiv:quant-ph/0608197; F. Verstraete, D. Porras, and J.I. Cirac,
Phys. Rev. Lett. \textbf{93}, 227205 (2004).

\bibitem{PEPS} F. Verstraete and J.I. Cirac, cond-mat/0407066; V. Murg, F. Verstaete, and J.I. Cirac,
 Phys. Rev. A \textbf{75}, 033605 (2007).

\bibitem{iTEBD} G. Vidal, Phys. Rev. Lett. \textbf{98}, 070201 (2007).

\bibitem{iPEPS} J. Jordan \textit{et al.}, arXiv:cond-mat/0703788.

\bibitem{PBC_TN} F. Verstraete and G. Vidal, in preparation.

\bibitem{TI} For simplicity, we assume that the system is invariant
under translations by one lattice site and that the ground state is
represented with a TN that consists of copies of the same tensor.
However, it is straightfoward to generalize the present discussion
to states that are invariant under translations by some finite
number of lattice sites, by considering a TN with the same symmetry.


\bibitem{multiple} The extension of our discussion to systems
depending on more than one control parameters is straightforward.

\bibitem{firstorder} Consider a quantum system described by a Hamiltonian
$H=H_0+\lambda H_1$,  with $[H_0,H_1]=0$. Suppose there is a
transition point $\lambda_c$ due to level crossing. Therefore,
ground states are the same (orthogonal) if they are in the same
(different) phase(s). This implies that $d(\lambda_1,\lambda_2)=1
(0)$ if $\lambda_1$ and $\lambda_2$ are in the same (different)
phase(s). That is, $d(\lambda_1,\lambda_2)$ is discontinuous at
$\lambda_1=\lambda_c \; (\lambda_2=\lambda_c)$. This argument may be
extended to a general model Hamiltonian with the above situation as
a special case.

\bibitem{book1} J. Zinn-Justin, \textit{Quantum Field Theory and Critical
Phenomena}, Oxford University Press, 1993, Oxford.

\bibitem{book2} R. Baxter, \textit{Exactly Solved Models in Statistical
Mechanics}, Academic Press, 1982, London.

\bibitem{classical} We emphasize that temperature is not a control parameter in the classical statistical
vertex lattice model.

\bibitem{1Dmapping} The mapping for one-dimensional quantum systems has been briefly discussed in the third reference in \cite{zhou}.

\bibitem{TI_PEPS} D. Perez-Garcia \textit{et al.}, arXiv:0707.2260.

\bibitem{non-degenerate} The case $|\mu_0|=|\mu_1|$ can also be dealt with by considering the corresponding eigenvectors.

\bibitem{previous} The calculation of the fidelity from a MPS was originally derived in the first reference in \cite{zhou}. The authors had previously communicated the result to P. Zanardi and it is also discussed in the third reference in \cite{fidelity}.


\bibitem{blote} H.W.J. Blote and Y. Deng,  Phys. Rev. E \textbf{66}, 066110 (2002).

\end{thebibliography}
\end{document}